\documentclass[conference]{IEEEtran}
\IEEEoverridecommandlockouts
\usepackage{cite}
\usepackage{amsmath,amssymb,amsfonts}
\usepackage{algorithmic}
\usepackage{graphicx}
\usepackage{textcomp}
\usepackage{xcolor}
\def\BibTeX{{\rm B\kern-.05em{\sc i\kern-.025em b}\kern-.08em
    T\kern-.1667em\lower.7ex\hbox{E}\kern-.125em}}
\begin{document}

\title{Using Collaborative Filtering to Recommend Champions in League of Legends}

\author{\IEEEauthorblockN{Tiffany D. Do\IEEEauthorrefmark{1}, Dylan S. Yu\IEEEauthorrefmark{2}, Salman Anwer\IEEEauthorrefmark{2}, and Seong Ioi Wang\IEEEauthorrefmark{2}}\IEEEauthorblockA{\IEEEauthorrefmark{1}College of Engineering and Computer Science\\University of Central Florida\\ Orlando, Florida \\Email: tiffanydo@knights.ucf.edu}\IEEEauthorblockA{\IEEEauthorrefmark{2}School of Engineering and Computer Science\\University of Texas at Dallas\\ Richardson, Texas\\Email: \{dylanyu461, salman.anwer.tr, seongiwang\}@gmail.com}}


\maketitle

\begin{abstract}
League of Legends (LoL), one of the most widely played computer games in the world, has over 140 playable characters known as champions that have highly varying play styles. However, there is not much work on providing champion recommendations to a player in LoL. In this paper, we propose that a recommendation system based on a collaborative filtering approach using singular value decomposition provides champion recommendations that players enjoy. We discuss the implementation behind our recommendation system and also evaluate the practicality of our system using a preliminary user study. Our results indicate that players significantly preferred recommendations from our system over random recommendations.
\end{abstract}


\section{Introduction}
League of Legends (LoL), originally released in 2009 by Riot Games, is one of the most widely played computer games in the world. There are an estimated 100 million active players each month as of 2016 \cite{volk_2016}. The game currently has over 140 playable characters, known as champions, from which players can choose from at the start of every match \cite{champlist}.
These champions have widely varied characteristics and abilities that allow a player to truly customize their way of playing LoL. With so many available champions, our aim was to create a recommendation system that would enable players to discover champions they might like based on their current champion preferences. 

In this paper, we implemented a champion recommendation system using collaborative filtering, specifically the singular value decomposition (SVD) algorithm. We conducted a preliminary user study to evaluate our system. Our results indicate that players significantly preferred our system's recommendations over random recommendations. 

\section{Related Work}
In the broader Multiplayer Online Battle Arena (MOBA) genre, of which LoL is firmly situated within, previous work on recommendation systems share one broad underlying assumption: that players are primarily motivated by maximizing win-rate. Under this assumption, previous MOBA recommendation systems provided recommendations that optimize either team champion synergy (draft-based recommenders) \cite{1hanke2017recommender, 3chen2018recommender,  5porokhnenko2019recommender, 6wang2016outcome, 7semenov2016outcome, 8summerville2016draft} or player item synergy (item-based recommenders) \cite{2smit2019recommender, 4looi2018recommender}. Our approach, however, does not follow this assumption and instead provides recommendations that focus on player enjoyment.

Multiple previous recommenders focused on maximizing win-rate through synergy. Smit \cite{2smit2019recommender} created a recommender system for items in LoL, allowing a dynamic approach to a recommender system that adapts to the game over defined phases, utilizing Artificial Neural Nets trained on game states and win-rate. 
Smit's work focused on item synergy instead of champion synergy, but it is still a notable work on a recommender oriented towards maximizing win-rate in MOBAs. With respect to the draft-based recommenders, Chen et al. \cite{3chen2018recommender} modeled the champion drafting process as a combinatorial game and utilized a Monte Carlo Tree Search with a Neural Net reward function to simulate and back-propagate dynamically optimal champion picks with respect to win-rate. While their approach achieved promising results over baseline strategies, their system targeted teams of players that are primarily concerned with the win-rate of their team's champion synergy, whereas we focus on individual player enjoyment. Their approach also did not take into account any player specific proclivities such as desired roles or favored champions. 

Our approach does not follow the underlying assumptions of the previous studies. The goal of our system is completely different from finding the optimal item or team. While previous works can be classified as player-win or team-win oriented, our work firmly targets player enjoyment. Our recommendations are centered on the following assumptions: 
1) a player is primarily driven to play champions based on their enjoyment of those champions, 
2) a player's current enjoyment of a champion is related to their historical engagement with that champion, and 
3) player engagement is a suitable driver for a general-audience MOBA champion recommender.
Our approach draws inspiration from Paterek's \cite{paterek2007netflix} substantially successful work on Netflix movie recommendations, which has similar problem characteristics to our own, given their focus on user ratings and the presence of high user engagement. 

\section{Champion Recommendation System}
\subsection{Collaborative Filtering}
We built our champion recommender based on a collaborative filtering approach \cite{Sarwar:2001}, specifically the SVD algorithm. LoL is suitable for collaborative filtering as there is a massive base of users with diverse tastes and high user engagement, thus avoiding the common issues of cold starts, rating sparsity, and user homogeneity.

We used an unbiased version of the SVD algorithm since we normalized the training data. The SVD algorithm was suitable as it provides low rank latent factor discovery and user-champion mapping onto this latent space. This helps mitigate popularity bias, which is a common issue in many recommender systems where popular items are frequently recommended over more niche items \cite{popbias}. Consequently, the SVD approach allows for closely tailored recommendations to a user's taste, which is important due to the diversity of champions available.

Our model maps both champions (i.e., items) and players (i.e., users) with a joint latent factor space of dimensionality $f$, such that user-item preferences are modeled as inner products of that space. Suppose each champion $i$ is associated with a vector $q_i \in \mathbb{R}^f$ and each player $u$ is associated with a vector $p_u \in \mathbb{R}^f$. For any champion $i$, $q_i$ represents the extent to which a champion $i$ possesses the corresponding factor. Similarly, for a user $u$, each row in $p_u$ represents $u$'s interest in the corresponding factor. We then follow the standard SVD algorithm described by Koren et al. \cite{koren2009matrix} to train our model using a Python implementation of the SVD algorithm from the Surprise library \cite{Surprise}.

\subsection{Building a training set}
SVD utilizes user ratings to build a model that can provide recommendations \cite{koren2009matrix}. While explicit ratings of how much a player enjoys a champion are not available, we are able to approximate their preferences by how often they have played a champion. Our recommendation system however, is further limited to the data made public by Riot Games. As Riot Games does not provide public access to how many times a player has played a champion, we instead make use of Champion Mastery Points $(C_{MP})$, a metric that is publicly accessible, which is an integer approximation of how often a player has played a certain champion. 
A player has a $C_{MP}$ for each champion in the game, where $0 \ \leq C_{MP} < \infty$.
Every time a player selects a  champion for a game, the player's $C_{MP}$ for that champion increases by an integer value depending on their performance. There is no way to decrease $C_{MP}$. However, due to these characteristics, a user that plays the game more frequently would have a higher overall $C_{MP}$ than a user who plays less frequently. To mitigate this issue, we normalized these scores to a linear scale between 1 and 100, such that a user's most played champion is rated as a 100, regardless of its raw $C_{MP}$ value. While players' $C_{MP}$ values generally follow a Poisson distribution, we found that a linear scaling performed the best in a pilot study, in part due to its mitigation of popularity bias. The use of $C_{MP}$ serves our aims well, as we believe that behaviorally observed preferences are more indicative of true preferences in this domain.

We pulled the $C_{MP}$ values of 2514 random active players (i.e., players who have played recently) using Riot’s official API \cite{riotapi} in order to train our model. We used this dataset of 2514 random players and their respective $C_{MP}$ values to build our recommender system. Our dataset and code is available at \cite{github}.

\subsection{Finding Recommendations}
To find recommendations for a player, we provided our system with a player's summoner name (their unique username) and queried Riot Games' API in order to find their top five champions sorted by $C_{MP}$. Table 1 displays an example of a sample player's top five $C_{MP}$ values and their normalized $C_{MP}$ ratings. For the queried user, we selected only the top five champions of a player because the sum total $C_{MP}$ of their top five champions typically account for a large majority of their sum total $C_{MP}$ among all champions. Furthermore, $C_{MP}$ values are not explicit ratings and a low $C_{MP}$ value does not necessarily mean that a player does not enjoy a champion. For example, a low $C_{MP}$ value could simply indicate that a player has not yet tried a champion very much. We used these five $C_{MP}$ values as the user's preferences to find ratings for all champions outside of their top five champions. After the system calculates ratings, it sorts these scores and outputs the top five respective champions as recommendations.
\begin{table}[htbp]
\caption{Example list of a player's top 5 champions and their $C_{MP}$ values}
\begin{center}
\begin{tabular}{| l | l | l |}
    \hline
    Champion & $C_{MP}$ & Normalized $C_{MP}$ \\ \hline
    Nami & 367,191 & 100\\ 
    Zyra & 136,709 & 38 \\
    Cassiopeia & 106,064 & 29 \\
    Janna & 89,306 & 25\\
    Lulu & 59,486 & 17\\
    \hline
\end{tabular}
\label{cmp-table}
\end{center}
\end{table}
\section{Survey Experiment}
We conducted a preliminary survey experiment to determine the practicality of our recommender system.

\subsection{Procedure}
Participants were asked for their summoner name which was used to generate recommendations. They were then given a survey that asked for their explicit champion ratings for 10 champions. 5 of these champions were our system's top recommendations and the other 5 were randomly selected from the remaining pool of champions. We chose to survey only 10 champions to keep the survey relatively short. The order of the champions presented was randomly determined to avoid introducing bias. We then analyzed the mean ratings given by the user and compared the recommendation ratings to the random ratings. Figure \ref{fig:survey} is an example of a survey given to a user. 
\begin{figure}
\begin{flushleft}
\begingroup
\setlength{\tabcolsep}{10pt}
\resizebox{\linewidth}{!}{%
\begin{tabular}{|@{}c@{}|}
    \hline
    \renewcommand{\arraystretch}{1.2} 
    \resizebox{\linewidth}{!}{%
    \begin{tabular}{l}
        How would you rate your enjoyment of each champion from 1-10? \\
        1 = I do not like playing this champion \\
        10 = This champion is very enjoyable to play
    \end{tabular}}\\
    \renewcommand{\arraystretch}{1.5} 
    \resizebox{\linewidth}{!}{%
    \begin{tabular}{l r l r}
         Karma & \rule{1cm}{1pt} &
         \hspace{2em}Graves *  & \rule{1cm}{1pt} \\
         Jax * & \rule{1cm}{1pt} &
         \hspace{2em}Sona & \rule{1cm}{1pt} \\
         Zilean & \rule{1cm}{1pt} &
         \hspace{2em}Syndra * & \rule{1cm}{1pt} \\
         Thresh & \rule{1cm}{1pt} &
         \hspace{2em}Warwick * & \rule{1cm}{1pt} \\
         Zed & \rule{1cm}{1pt} &
         \hspace{2em}Draven * & \rule{1cm}{1pt} \\
    \end{tabular}}\\
    \hline
\end{tabular}}
\endgroup
\end{flushleft}
    \caption{Sample survey given to a user where * indicates a random recommendation. This is not visible to the subject}
    \label{fig:survey}
\end{figure}

\subsection{Participants}
We recruited 30 unpaid volunteers (24 male, 6 female) from the university club LOLUTD (League of Legends at The University of Texas at Dallas). Participants were active and experienced players. 

\subsection{Research Questions and Hypotheses}
\textbf{RQ1: Do players enjoy our system's recommendations more than random recommendations?}

\textit{H1: Players will enjoy our system's recommendations more than random recommendations.} We expect that players will rate system recommendations higher than random recommendations.

\subsection{Results}
\begin{figure}[ht]
  \centering
  \includegraphics[width=\linewidth]{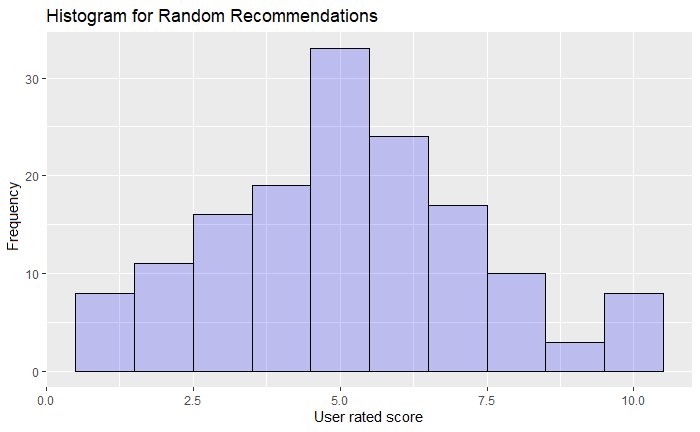}
  \caption{Histogram of user rated scores given to random recommendations}
  \label{fig:randomhist}
  \vspace{1em}
  \includegraphics[width=\linewidth]{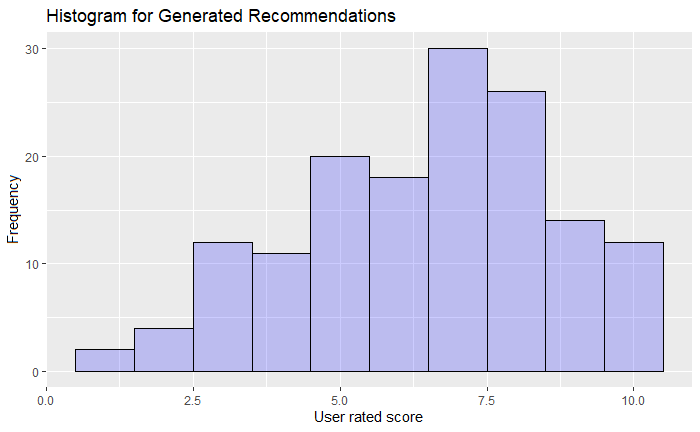}
  \caption{Histogram of user rated scores given to system generated recommendations}
  \label{fig:systemhist}
\end{figure}

Since we have complete access to all survey data points, we tested for significance using a one-sided two sample Z-test for means. System recommendations had a mean of 6.46 and random recommendations had a mean of 5.18. Given $p=0.01257$, we reject the null hypothesis $H_0$ that the means are equal at the $\alpha=0.05$ level and can conclude that system recommendations are rated significantly higher. We can tentatively state that users prefer the system generated recommendations more than random recommendations. This preference is evident in the histograms of scores as seen in Figures \ref{fig:randomhist} and \ref{fig:systemhist}. In Figure \ref{fig:randomhist}, the distribution of user rated scores seems relatively normal, which is to be expected of random recommendations. In Figure \ref{fig:systemhist}, the distribution of user rated scores is skewed to the left, showing that users rate system recommendations fairly high. Therefore, our results support H1.

\section{Discussion}
Our preliminary results indicate that an SVD-based collaborative filtering approach can recommend champions that players enjoy significantly more than random recommendations. This system could help introduce players to champions that they may have otherwise overlooked. Given the considerable number of champions in LoL, this can be a useful service to drive player engagement. Furthermore, our results show that $C_{MP}$ is a useful measure for player champion preference. These findings can help game designers and analysts identify player groups and preferences using $C_{MP}$. Additionally, our approach emphasizes recommending enjoyable champions to players rather than recommending champions that will increase a player's win rate. 

\subsection{Exploration of other algorithms}
We also built models using the Slope One \cite{lemire2005slope} and SVD++ algorithms in the Surprise library and obtained lackluster results.
We found Slope One to be highly susceptible to popularity bias; popular champions such as Ezreal, Lee Sin, and Yasuo were consistently recommended for users regardless of user preferences. 
While theoretically more accurate, the SVD++ model had a prohibitive run time for the aims of our project with only minor improvement in practical accuracy. We ran a test using an Intel i7-6700 with 32GB of RAM on our dataset of 311,727 rows. A single run using SVD++ in this environment took over an hour, which was not feasible for our application where users would want results in a reasonable timeframe. 

\subsection{Limitations}
Metagames in LoL are what players perceive to be the best strategy at some given time in the game \cite{donaldson2017mechanics}. These different metagames skew the popularity of certain champions based on their strength in a specific metagame. While this does affect our system, we believe that this effect balances out due to both the evolving metagames that give every champion their time in the limelight and people's natural preference to play a champion that they enjoy, even though it may impact their chances of winning.

Another consideration is that champions may change considerably due to champion reworks. Reworks may change a champion's skills and appearance, but are infrequently done to old champions that are in need of an update \cite{champUpdate}. As $C_{MP}$ is maintained regardless of any changes to the champions, our model will be misled for some user preferences. Additionally, new champions are occasionally added to the game. Older champions will have higher average $C_{MP}$ in comparison and newer champions will be less likely to be recommended. 

It is important to note that our survey is preliminary and was conducted to assess the potential of our recommendation system. Most participants were active, experienced players, while our system should ideally be suitable for the general, amateur player base. Additionally, our participants were also all college students and were mostly male.

\subsection{Conclusion and Future Work}

In this paper, we implemented a champion recommendation system for LoL using an SVD-based collaborative filtering approach. Our system uses champion mastery points to indicate player preferences. We conducted a preliminary user study that validates the practicality of our system and indicates that users enjoy our system's recommendations significantly more than random recommendations.

In the future, we plan to rerun our survey study using optimal parameters. The recommender tested in our user study made use of the naive SVD model in the Surprise library. 
The default parameters are \cite{Surprise}: $epoch = 20$, $\lambda = 0.005$, and $\gamma = 0.02$, where $\lambda$ is the regularization term and $\gamma$ is the learning rate.
The results obtained in our preliminary study indicated that the SVD model is appropriate for a champion recommender system. Following this validation, we tuned these model parameters using a grid search to maximize theoretical accuracy by minimizing root mean square error \cite{Surprise}. The optimal parameters found are the following: $epoch = 20$, $\lambda=0.4$, and $\gamma = 0.0005$.
By using these parameters for model fitting, stronger results in a user study would likely be obtained. Furthermore, to better tailor our system to individual player profiles, we could dynamically adjust the number of reference champions used for generating recommendations. This would allow for recommendations that meet the needs of players whose $C_{MP}$ values are more uniformly distributed.

We also plan to run a more comprehensive survey study. We could compare our system recommendations to more naive recommendation systems, such as a selection of the most played champions. Additionally, we could follow a more standardized methodology and also survey a more diverse and representative population. 

\section{Acknowledgments}
This project was initially conceived in a graduate machine learning course at The University of Texas at Dallas. The authors would like to thank Dr. Anjum Chida and Dr. Cuneyt Akcora for their valuable comments and teachings.

\bibliographystyle{IEEEtran}
\bibliography{IEEEexample}

\begin{thebibliography}{10}
\providecommand{\url}[1]{#1}
\csname url@samestyle\endcsname
\providecommand{\newblock}{\relax}
\providecommand{\bibinfo}[2]{#2}
\providecommand{\BIBentrySTDinterwordspacing}{\spaceskip=0pt\relax}
\providecommand{\BIBentryALTinterwordstretchfactor}{4}
\providecommand{\BIBentryALTinterwordspacing}{\spaceskip=\fontdimen2\font plus
\BIBentryALTinterwordstretchfactor\fontdimen3\font minus
  \fontdimen4\font\relax}
\providecommand{\BIBforeignlanguage}[2]{{%
\expandafter\ifx\csname l@#1\endcsname\relax
\typeout{** WARNING: IEEEtran.bst: No hyphenation pattern has been}%
\typeout{** loaded for the language `#1'. Using the pattern for}%
\typeout{** the default language instead.}%
\else
\language=\csname l@#1\endcsname
\fi
#2}}
\providecommand{\BIBdecl}{\relax}
\BIBdecl

\bibitem{volk_2016}
\BIBentryALTinterwordspacing
P.~Volk, ``League of legends now boasts over 100 million monthly active players
  worldwide,'' Sep 2016. [Online]. Available: \url{https://goo.gl/iCo4sa}
\BIBentrySTDinterwordspacing

\bibitem{champlist}
\BIBentryALTinterwordspacing
``Champions.'' [Online]. Available:
  \url{https://na.leagueoflegends.com/en/game-info/champions/}
\BIBentrySTDinterwordspacing

\bibitem{1hanke2017recommender}
L.~Hanke and L.~Chaimowicz, ``A recommender system for hero line-ups in moba
  games,'' in \emph{Thirteenth Artificial Intelligence and Interactive Digital
  Entertainment Conference}, 2017.

\bibitem{3chen2018recommender}
Z.~Chen, T.-H.~D. Nguyen, Y.~Xu, C.~Amato, S.~Cooper, Y.~Sun, and M.~S.
  El-Nasr, ``The art of drafting: a team-oriented hero recommendation system
  for multiplayer online battle arena games,'' in \emph{Proceedings of the 12th
  ACM Conference on Recommender Systems}, 2018, pp. 200--208.

\bibitem{5porokhnenko2019recommender}
I.~Porokhnenko, P.~Polezhaev, and A.~Shukhman, ``Machine learning approaches to
  choose heroes in dota 2,'' in \emph{2019 24th Conference of Open Innovations
  Association (FRUCT)}.\hskip 1em plus 0.5em minus 0.4em\relax IEEE, 2019, pp.
  345--350.

\bibitem{6wang2016outcome}
W.~Wang, ``Predicting multiplayer online battle arena (moba) game outcome based
  on hero draft data,'' Ph.D. dissertation, Dublin, National College of
  Ireland, 2016.

\bibitem{7semenov2016outcome}
A.~Semenov, P.~Romov, S.~Korolev, D.~Yashkov, and K.~Neklyudov, ``Performance
  of machine learning algorithms in predicting game outcome from drafts in dota
  2,'' vol. 661, 04 2016.

\bibitem{8summerville2016draft}
A.~Summerville, M.~Cook, and B.~Steenhuisen, ``Draft-analysis of the ancients:
  predicting draft picks in dota 2 using machine learning,'' in \emph{Twelfth
  Artificial Intelligence and Interactive Digital Entertainment Conference},
  2016.

\bibitem{2smit2019recommender}
R.~Smit, ``A machine learning approach for recommending items in league of
  legends,'' 2019.

\bibitem{4looi2018recommender}
W.~Looi, M.~Dhaliwal, R.~Alhajj, and J.~Rokne, ``Recommender system for items
  in dota 2,'' \emph{IEEE Transactions on Games}, 2018.

\bibitem{paterek2007netflix}
A.~Paterek, ``Improving regularized singular value decomposition for
  collaborative filtering,'' in \emph{Proceedings of KDD cup and workshop},
  vol. 2007, 2007, pp. 5--8.

\bibitem{Sarwar:2001}
\BIBentryALTinterwordspacing
B.~Sarwar, G.~Karypis, J.~Konstan, and J.~Riedl, ``Item-based collaborative
  filtering recommendation algorithms,'' in \emph{Proceedings of the 10th
  International Conference on World Wide Web}, ser. WWW '01.\hskip 1em plus
  0.5em minus 0.4em\relax New York, NY, USA: ACM, 2001, pp. 285--295. [Online].
  Available: \url{http://doi.acm.org/10.1145/371920.372071}
\BIBentrySTDinterwordspacing

\bibitem{popbias}
H.~Abdollahpouri, ``Popularity bias in ranking and recommendation,'' 01 2019.

\bibitem{koren2009matrix}
Y.~Koren, R.~Bell, and C.~Volinsky, ``Matrix factorization techniques for
  recommender systems,'' \emph{Computer}, no.~8, pp. 30--37, 2009.

\bibitem{Surprise}
N.~Hug, ``{S}urprise, a {P}ython library for recommender systems,''
  \url{http://surpriselib.com}, 2017.

\bibitem{riotapi}
\BIBentryALTinterwordspacing
{Riot Games}, ``Riot developer api.'' [Online]. Available:
  \url{https://developer.riotgames.com/api-methods/}
\BIBentrySTDinterwordspacing

\bibitem{github}
\BIBentryALTinterwordspacing
D.~Yu, T.~Do, and S.~Anwer, ``Project repository.'' [Online]. Available:
  \url{https://github.com/JonahFarc/League-Champion-Recommender}
\BIBentrySTDinterwordspacing

\bibitem{lemire2005slope}
D.~Lemire and A.~Maclachlan, ``Slope one predictors for online rating-based
  collaborative filtering,'' in \emph{Proceedings of the 2005 SIAM
  International Conference on Data Mining}.\hskip 1em plus 0.5em minus
  0.4em\relax SIAM, 2005, pp. 471--475.

\bibitem{donaldson2017mechanics}
S.~Donaldson, ``Mechanics and metagame: Exploring binary expertise in league of
  legends,'' \emph{Games and Culture}, vol.~12, no.~5, pp. 426--444, 2017.

\bibitem{champUpdate}
\BIBentryALTinterwordspacing
``Champion updates,'' 2018. [Online]. Available:
  \url{https://support-leagueoflegends.riotgames.com/hc/en-us/articles/202294884-Champion-Update-Schedule}
\BIBentrySTDinterwordspacing

\end{thebibliography}
\end{document}